\documentstyle[aps,multicol,prl,epsf]{revtex}

\begin{document}
\tighten
\draft

\title{Effect of a Variable-Spin Quantum Dot on a Wire Conductance}

\author{Maria A. Davidovich$^1$, E.V. Anda$^1$, C.A. B\"usser$^2$, and G. Chiappe$^2$ \\
\it $^1$Departamento de F\'{\i}sica, Pontificia Universidade 
Cat\'olica do Rio de Janeiro, C.P. 38071-970, Rio de Janeiro, RJ, Brazil \\
\it $^2$Departamento de F\'{\i}sica,
Facultad de Ciencias Exactas y Naturales, \it Universidad de Buenos Aires
Ciudad Universitaria, 1428, Buenos Aires, Argentina}
\maketitle

\begin{abstract}
A numerically exact calculation of the $T=0$ transport properties of a quantum wire interacting with a lateral two-level quantum dot is presented. The wire conductance is calculated for all different states of charge and spin of the quantum dot. For a dot with two electrons we obtain an enhancement of the Kondo temperature at the singlet-triplet transition and a non-universal scaling law for its dependence upon the dot energy spacing. We find that the  Kondo correlation is stronger for a dot spin $S_D \sim 1$ than for $S_D \sim 1/2$. In both cases the wire current is totally quenched by the Kondo effect. When the dot is in the mixed-valence regime and $1/2 < S_D < 1$ the wire conductance is partially quenched except in a very small region of gate potential where it  reaches the maximum value $e^2/h$. 
 \end{abstract}

\begin{multicols}{2}
In recent years an important experimental and theoretical effort has been dedicated to the understanding of the physical properties of a system constituted by a dot connected to two leads under the effect of an external potential\cite{variosT,variosT1,variosE,Alfredo,Goldhaber,Georges}. This system has shown to be a versatile device to study the Kondo effect for a variety of regimes that can be accessed by changing continuously the values of the parameters that define the system. For temperatures below $T_K$, in the Kondo regime, the electrons circulate along the system through a channel at the Fermi level that disappears as the temperature is increased. 

Recently, in a very interesting experiment\cite{Kouven}, the measurement of the transport through a dot with energy levels tuned by the application of an external magnetic field has demonstrated that the current through an even-charged dot connected to two leads is controlled by the Kondo correlation between the $S_D=1$ spin of the dot triplet state and the conduction electron spins. This correlation gets stronger as the magnetic field energy splitting is increased, and reaches its maximum value when the system is in the singlet-triplet transition, beyond which the Kondo effect dies out very rapidly. 

This problem was theoretically studied by using poor's man scaling arguments and the mapping of a generic model of a dot undergoing a singlet-triplet transition onto the two impurity Kondo model\cite{Glazman,Eto}. These theoretical works restrict their attention to the middle of the Coulomb blockade valley, situation where the dot has an integer well defined number of electrons, so that all mixed valence regimes are eliminated. Assuming that there is an even number of electrons in the dot and that the system is in the immediate vicinity of the singlet-triplet transition they show that the Kondo effect is enhanced by the competition between the triplet and the singlet states. 

In this letter we report the results of a numerically exact solution for a two-level quantum dot interacting with a wire, at $T=0$, for the whole range of parameters that permits the investigation of all possible regimes and states of charge and spin of the system. By varying the gate potential applied to the dot  and the dot level energy spacing we are able to study the wire conductance for the dot in the mixed valence regime, in the triplet $(S=1)$, in the doublet $(S=1/2)$ and in the singlet $(S=0)$ spin states.

The system we consider consists of a dot laterally connected to a conducting wire. This is a very interesting configuration as it mimics the situation of a metal doped by magnetic impurities, where the localized spins belong to an electronic state outside the conduction band\cite{Coqblin}. In this configuration, as shown in a recent theoretical work\cite{Ereferee} restricted to a one level dot, the Kondo correlation among the dot spin $S_D=1/2$ and its neighbor conducting spins provides a scattering mechanism for the electrons that quenches the circulating current. This behavior was explained by assuming that the system is a Fermi liquid that satisfies the Friedel-Langreth sum rule and by using the mean field slave boson infinite Coulomb repulsion approximation. 

In our calculations we use parameter values that are compatible with the experimental situation. In particular, as we take the Coulomb interaction to be finite we have the appropriate energy scale for the Kondo temperature. Since we use a Lanczos diagonalization, our analysis is restricted  to $T=0$.
For the dot with a fixed number of two electrons and total spin $S_D \sim 1$, we obtain an enhancement of the Kondo effect when the dot energy spacing is equal to the exchange interaction, at the singlet-triplet transition. In this case, the wire conductance is cancelled due to strong interference effects. As the number of particles and spin of the dot is varied, the conductance goes from zero to $e^2/h$, according to the different regimes of the dot. We find that the $S=1$ Kondo correlation is stronger than the usual $S=1/2$. Our results are compatible with experimental observations on a system where the current goes through the dot\cite{Kouven}.

In order to be able to incorporate the physics associated with the singlet-triplet transition the dot is described by two levels. Contrary to the poor's man scaling analysis, we allow it to have different states of charge, as it occurs in any experiment where the gate potential is varied. Part of the system, consisting of a cluster which includes the two-level dot, is exactly solved using a Lanczos algorithm\cite{Dagotto} and then embedded into the rest of the wire. We are able to show that, for reasonable reduced cluster sizes, due to the way in which the embedding is done, the results are independent of the cluster size\cite{nosso} and, as a consequence, numerically exact. We calculate the state of charge and spin at the dot, the density of states projected onto the QD, and the current circulating through the wire, as a function of the gate potential  applied to the QD and of its energy level energy spacing.

The system is represented by an Anderson impurity first-neighbor tight-binding Hamiltonian. In a cluster of $2(M+1)$ sites, (the number is taken to be even to maintain the cluster symmetry), the two states of the QD are denoted by $\alpha$  and $\beta$ and the other $2M+1$ sites belonging to the wire numbered from $-M$ to $M$. The total Hamiltonian can be written as, 
\begin{eqnarray}
H&=&\sum_{r=\alpha,\beta\atop \sigma}(V_g+\epsilon_r)n_{r\sigma}+{U\over 2}\sum_{r=\alpha,\beta\atop \sigma} n_{r\sigma}n_{r\bar\sigma}\nonumber\\
&+&U\sum_{\sigma\sigma'} n_{\alpha\sigma}n_{\beta\sigma'}+ J(S_\alpha S_\beta)+t'\sum_\sigma[(c_{\alpha\sigma}+c_{\beta\sigma})c^+_{0\sigma}\nonumber\\
&+& c.c.]+\sum_{i,\sigma}t_{ij}c_{i\sigma}^+c_{j\sigma}
\end{eqnarray}
where $U$, $J$ and $V_g$ are, respectively, the Coulomb repulsion in the dot, the exchange energy and the gate potential applied to the two dot states with energies $\epsilon_\alpha$ and $\epsilon_\beta$. The hopping matrix elements between these two states and their first neighbor, site 0 at the wire, taken for simplicity to be equal, are given by $t'$, and the $t_{ij}=t$ are the nearest-neighbor hopping elements within the wire.

The Fermi energy is taken to be $\epsilon_F=0$. The dot levels, with energy spacing $\Delta=\epsilon_\alpha-\epsilon_\beta$, interact through the exchange coupling and can be changed continuously by varying the gate potential applied to the QD.

To obtain the properties of the system we calculate the one particle Green functions $G_{ij}$. They are made to satisfy a Dyson equation $\hat G=\hat g+\hat g \hat T \hat G$ where $\hat g$ is the cluster Green function matrix and $\hat T$ is the matrix of the coupling Hamiltonian between the cluster and the rest of the system. The undressed Green function $\hat g$ is calculated using the cluster ground state obtained by the Lanczos method. In order to guarantee consistency the charge of the dressed and undressed cluster is imposed to be the same. We calculate $\hat g$ as a combination  of the Green function of n and $n+1$ electrons with weight $1-p$ and $p$, $\hat g = (1-p)\hat g_n + p \hat g_{n+1}$. The charge of the undressed cluster is $q_c=(1-p)n + p(n+1)$\cite{Valeria}. The charge of the cluster when linked to the leads can be expressed  as $Q_c=2\int^{\epsilon_F}_{-\infty}\sum_i Im\,Gii(\omega)d\omega$,
where $i$ runs over all the cluster sites. This equation together with the imposed condition $q_c=Q_c$ constitutes a system of two equations which requires a self-consistent solution to obtain $p$ and $n$. Using the Keldysh\cite{Keldysh} formalism the conductance can be written as              
\begin{equation}
G={e^2t^2 \over h}|G_{00}|^2[\rho(\epsilon_F)]^2
\end{equation}
where $G_{00}$ is the Green function of the wire site connected to the dot, while $\rho(\epsilon_F)$is the density of states at the Fermi level at the first neighbors of site $0$, when disconnected from it.

We discuss first the transport properties in the wire as a function of the QD  energy spacing $\Delta$, maintaining fixed the number of electrons at the dot; then, for  fixed value of $\Delta$ we vary the number of particles at the QD, by changing  $V_g$. All energies are in units of $U$. 
We take for the parameters that define the system  values compatible with experiments\cite{Goldhaber}, $t'=0.6$ and ${\Gamma}={t'^{2}\over W}=0.08$, where $W$ is the wire bandwidth. 

In order to analyze the singlet-triplet transition we fix the gate potential such that the dot is charged with two particles. The results for the current in the wire and total spin at the dot are presented in Fig.1 as a function of $\Delta \over J$, the ratio  between the dot energy level spacing and the exchange coupling. As $\Delta$ increases the QD spin goes from a triplet to a singlet state. The conductance is a smooth function which interpolates between the value for a perfect insulator, when $S_D \sim 1$, and the value for a perfect conductor, when $S_D \sim 0$. This result reflects the effect of the $S=1$ Kondo correlation between the spin at the dot and the conduction electron spins in the wire, giving rise to a strong scattering that results in the cancellation of the wire conductance. An abrupt change of the current occurs around the singlet-triplet transition, ${\Delta \over J} \sim 1$, when the singlet and triplet states are degenerate. 
This result is the counterpart obtained by the poor's man scaling calculation in the case of a system in which the current goes through the dot itself\cite{Glazman,Eto}.

The study of the temperature dependence of the conductance would require a  huge computational effort, outside our capabilities. However, it is possible to obtain the Kondo temperature as a function of $\Delta$ from the density of states of the system. It is known that the Kondo contribution to the Green function in the immediate vicinity of the Fermi level assumes a Lorentzian form\cite{Hew} ${Z(T_k)\over E-\epsilon_F+iT_k}$,where the Kondo weight $Z(T_k)$ is a function of the Kondo temperature. From the analysis of the widths of the Kondo peaks for different dot energy splittings we obtain a Kondo temperature that increases with $\Delta$, has its maximum at the singlet-triplet transition, and decreases very abruptly for ${\Delta \over J} > 1$, as shown in Fig.2. This is compatible with the results for the conductance through a dot embedded in a wire as a function of the magnetic field energy splitting\cite{Kouven} where the Kondo effect is strong and only clearly observed near the singlet-triplet transition. 

For the dot embedded in a wire it has been predicted\cite{Glazman,Eto} in the limit $T_k(0)<< \delta$,  where $\delta = J-\Delta$, a power law dependence of the Kondo temperature on $\delta$ 
 \begin{equation}
T_K(\delta) =  T_K(0)\left[{T_K(0)\over \delta}\right]^\gamma 
\end{equation}

 In the inset of Fig.2 we show this relation in a logarithmic scale from which we obtain an exponent $\gamma=0.19$, inside the range of values established in Ref. $9$. As in that work, we have found that this  value is not universal and depends on $\Gamma$. As far as the character of the power law is concerned we coincide with these  authors. However, our results are numerically exact and do not require the truncation of the Hilbert space of the quantum dot, assumed in the application of the poor man's scaling.

With the purpose of studying other experimentally accessible regimes we fix the dot level energy spacing and calculate the current in the wire for all possible states of charge and spin of the dot by varying continuously the number of electrons in the QD. In this way we are able to study the mixed valence regime  and the evolution of the spin state of the QD from the doublet $(S=1/2)$, when the dot has an odd number of  electrons, to the triplet $(S=1)$ or the singlet $(S=0)$, when the number of electrons in the dot is even.

The results as a function of $V_g$ are displayed in Fig. 3, for  ${\Delta \over J} = 1$. The conductance in the wire and the total spin and charge at the QD are shown in Fig.3(a) and (b); the Kondo correlation of the dot total spin $S_D$ with the spin of the conduction electron at the neighboring site, $\langle\vec S_{D} \vec S_{c}\rangle$, and the spin correlation of the two states of the QD, $\langle\vec S_{\alpha} \vec S_{\beta}\rangle$ are shown in Fig.3(c). Since we choose the lowest dot energy $\epsilon_\alpha = 0$, for $V_g > 0$ both dot levels are above the Fermi level and the charge at the dot is about zero. As $V_g$ reduces, charge enters into the dot and the system goes through a charge fluctuation regime up to $V_g \sim -0.4$ where the dot has about one electron and total spin $S_D \sim 0.4$. At this point the Kondo correlation has a local maximum. This is the usual $S = 1/2$ Kondo state, which interferes with the wire current leading to strong scattering that results in the cancellation of the conductance.
As $V_g$ continues to decrease the system goes again into a mixed valence regime - the lowest level rearrange its charge between spin up and down and extra charge enters in the second dot level. The Kondo correlation diminishes and has a minimum around $V_g = -0.75$, where the full current circulating along the wire is restored. As charge keeps entering into the dot the Kondo correlation begins to increase again, and reaches an  absolute maximum value at $V_g \sim -1.9$, when there are just two electrons at the dot and the singlet and triplet states are degenerate. At this point the total dot spin and the dot spin-spin correlation $\langle\vec S_{\alpha} \vec S_{\beta}\rangle$ have maximum values, and the current is again cancelled. The $S=1$ Kondo regime is responsible for this strong scattering effect. 
As reflected on the curve for conductance as a function of $V_g$ the system has electron-hole symmetry with respect to $V_g=-1.9$. 

Notice the strong correlation between the variation of the current and of the Kondo spin correlation as a function of $V_g$. The bigger is the Kondo correlation the stronger are the interference effects which cause the degradation of the current. Moreover, as can be seen in Fig.3c, the Kondo correlation is stronger for the dot with two electrons and maximum total spin $S_D \sim 1$, in the singlet-triplet transition, than for the usual case of $S_D \sim 1/2$. This result is compatible with the measurements\cite{Kouven} of the conductance as a function of gate voltage in a system where the dot is embedded in a wire, where the $S=1/2$ Kondo effect is much weaker, and almost not observable, than the $S=1$ Kondo effect in the singlet-triplet transition. Notice also the way the charge enters into the dot, continuously rather than in steps as in the Coulomb blockade regime, reflecting the gradual way the Kondo peak move through the Fermi level as $V_g$ is varied.

In summary, we present numerically exact results for the transport properties of a system consisting of a two-level quantum dot laterally coupled to a wire, where the dot energy spacing is varied in order to mimic the effect of a magnetic field. In the special case where the dot has an integer number of electrons equal to two, the wire conductance shows an abrupt transition from a perfect insulating to a perfect conductor for the value of the energy spacing such that the dot singlet and triplet spin states are nearly degenerate.
Our results show a maximum Kondo temperature at this singlet-triplet transition and a power law dependence of $T_K$ with the dot energy splitting.
We have also explored the intermediate valence regime by varying the gate potential applied to the dot. In this way we were able to study the influence of the different states of the dot charge and spin upon the conductance in the wire. An analysis of the spin correlation permits to understand the influence of the Kondo state on the wire current. The conductance, in the range where the number of dot electrons goes from one to three, follows the same behavior as the Kondo correlation, being minimum when the Kondo correlation is maximum and vice-versa.
Our results are exact and compatible with the experiments in a system where the dot is inserted in the wire. We hope that transport measurements could be done in a wire with the dot in a lateral configuration, as discussed in this work.

We acknowledge CNPq, CAPES, FAPERJ, Antorchas/Vitae/Andes grant B-11487/9B003, CONICET and Fundacion Antorchas for financial support.

\begin{figure}
\epsfxsize=2.5in
\centerline{\epsffile{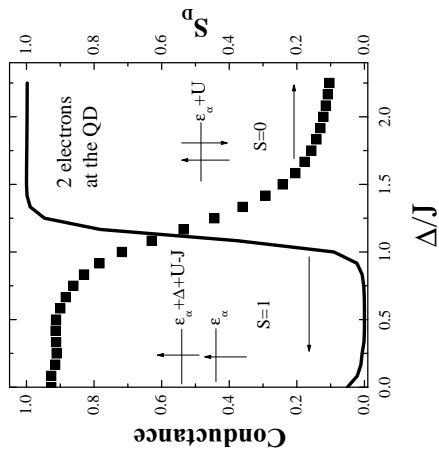}}
\epsfysize=2.5in 
\narrowtext
\caption{Conductance in units of ${e^2}/h$ (continuous line) and QD total spin (squares) as a function of $\Delta \over J$, for two electrons at the dot. The singlet-triplet transition corresponds to ${\Delta \over J} =1$.}
\label{1}
\end{figure} 

\begin{figure}
\epsfxsize=2.5in
\epsfysize=2.8in 
\centerline{\epsfbox{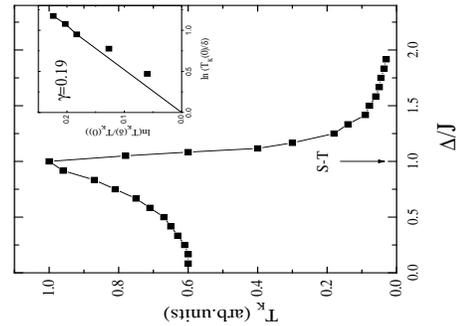}}
\narrowtext
\caption{Kondo temperature as a function of $\Delta \over J$. In the inset the exponent $\gamma$ is obtained from the logarithm of the equation $ T_K(\delta)= T_K(0)[{T_K(0)\over \delta}]^\gamma$.} 
\label{2}
\end{figure}

\begin{figure}
\epsfxsize=2.5in
\centerline{\epsffile{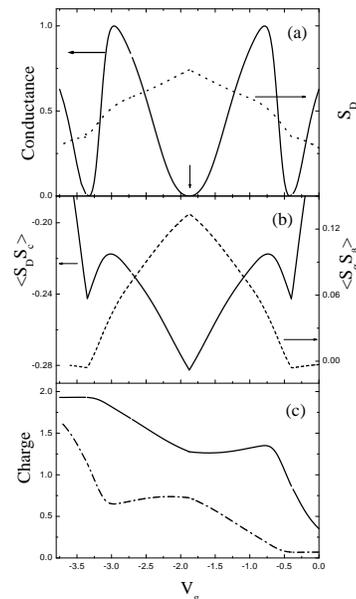}}
\epsfysize=3.5in
\narrowtext
\caption{Conductance, spin, charge and spin correlation as a function of $V_g$, for  ${\Delta \over J} =1$. (a) Conductance (continuous line) in units of ${e^2}/h$ and QD total spin (dotted line);(b) Charge in state $\alpha$ (continuous line) and in state $\beta$ (dashed-dotted line) in units of the electronic charge; (c) Kondo spin correlation, $\langle\vec S_{D} \vec S_{c}\rangle$ (continuous line) and spin correlation of the two states of the QD, $\langle\vec S_{\alpha} \vec S_{\beta}\rangle$ (dashed line).}  
\label{3}
\end{figure} 

\end{multicols}

\begin{references}

\bibitem{variosT} 
L.I. Glazman {\it et al.}, Europhys. Lett. {\bf 19} 623 (1992).

\bibitem{variosT1}
Y. Meir {\it et al.}, Phys. Rev. Lett. {\bf 66}, 3048 (1991); 
A. Groshev {\it et al.}, Phys. Rev. Lett. {\bf 66}, 1082 (1991).


\bibitem{variosE} 
L.P. Kowenhoven {\it et al.}, {\ it "Mesoscopic Electron Transport"}; L. Shon {\it et al.}, 
(eds.) NATO ASI Ser. E (1997); 
R. Ashoori, Nature {\bf 379}, 413 (1996).


\bibitem{Alfredo} 
T.K. Ng {\it et al.}, Phys. Rev. Lett. {\bf 61}, 1768 (1988);
L.I. Glazman {\it et al.}, JETP Lett. {\bf 47}, 452 (1988);
S. Hershfield {\it et al.}, Phys. Rev. Lett. {\bf 67}, 3720 (1991);
Y.Meir {\it et al.}, Phys. Rev. Lett. {\bf 70}, 2601 (1993); 
N.S. Wingreen {\it et al.}, Phys. Rev. B {\bf 49}, 11040 (1994); 
M.A. Davidovich {\it et al.}, Phys. Rev B {\bf 55}, R7335 (1997);
A. Levy Yeyati {\it et al.}, Phys. Rev. Lett. {\bf 83}, 600 (1999). 

\bibitem{Goldhaber} 
D. Goldhaber-Gordon {\it et al.}, Nature {\bf 391}, 156 (1998);
D. Goldhaber-Gordon {\it et al.}, {\it "Proceedings of the 24th International 
Conference on The Physics of Semiconductors"}; D. Gershoni {\it et al.}, (Eds.)
World Scientific, Singapure (1999);
S.M. Cronenwatt {\it et al.}, Science {\bf 281}, 540 (1998).

\bibitem{Georges} 
T. Aono {\it et al.}, J. Phys. Soc. Jpn. {\bf 67}, 1860 (1998);
A. Georges {\it et al.}, Phys. Rev. Lett. {\bf 82}, 3508 (1999).

\bibitem{Kouven}
S.Sasaki {\it et al.}, Nature {\bf 405}, 764 (2000).

\bibitem{Glazman}
M.Pustilnik and L.I.Glazman, Phys. Rev. Lett. {\bf 85}, 2993 (2000). 

\bibitem{Eto}
M.Eto and Y.V.Nazarov, Phys. Rev. Lett. {\bf 85}, 1306 (2000). 


\bibitem{Coqblin} 
B. Coqblin {\it et al.}, J. Phys. Soc. Jpn. {\bf 65} Suppl. B 64 (1996).

\bibitem{Ereferee} 
K. Kang {\it et al.}, Preprint, Cond-Matt. 0009235 (2001).


\bibitem{Dagotto} 
E. Dagotto. Rev. Mod. Phys. {\bf 66}, 763 (1994).


\bibitem{nosso}
C.A.Busser {\it et al.}, Phys. Rev B {\bf 62}, 9907 (2000)

\bibitem{Valeria} 
V. Ferrari {\it et al.}, Phys. Rev. Lett. {\bf 82}, 5088 (1999).

\bibitem{Keldysh} L.V. Keldysh, Sov. Phys. JETP {\bf 20}, 1018, (1965).

\bibitem{Hew} A.C.Hewson {\it "The Kondo Problem to Heavy 
Fermions"} Cambridge University Press, UK (1993).


\end{references}
\end{document}